\documentclass[iop, apj]{emulateapj}

\usepackage[cp1251]{inputenc}
\usepackage[english]{babel}
\RequirePackage[T2A]{fontenc}
\usepackage{amsmath}
\usepackage{amsfonts}
\usepackage{amssymb}
\usepackage{graphicx}

\shorttitle{Initiation and early evolution of the CME from EUV and white-light observations}
\shortauthors{A.~Reva, A.~Ulyanov et al.}

\begin{document}
\title{Initiation and early evolution of the Coronal Mass Ejection on May 13, 2009 from EUV and white-light observations}
\author{Reva A.A., Ulyanov A.S., Bogachev S.V., Kuzin S.V.}
\affil{Lebedev Physical Institute, Russian Academy of Sciences}
\email{reva.antoine@gmail.com}

\begin{abstract}
We present the results of the observations of a coronal mass ejection (CME), which occurred on May 13, 2009. The most important feature of these observations is that the CME was observed from the very early stage (the solar surface) up to a distance of 15 solar radii ($R_\odot$). Below 2~$R_\odot$, we used the data from the TESIS EUV telescopes obtained in the Fe 171~\AA\ and He 304~\AA\ lines, and above 2~$R_\odot$, we used the observations of the LASCO C2 and C3 coronagraphs. The CME was formed at a distance of 0.2–-0.5~$R_\odot$ from the Sun’s surface as a U-shaped structure, which was observed both in the 171~\AA\ images and in white-light.  Observations in the He~304~\AA\ line showed that the CME was associated with an erupting prominence, which was located not above---as predicts the standard model---but in the lowest part of the U-shaped structure close to the magnetic X-point. The prominence location can be explained with the CME breakout model. Estimates showed that CME mass increased with time. The CME trajectory was curved---its helio-latitude decreased with time. The CME started at latitude of 50$^\circ$ and reached the ecliptic plane at distances of 2.5~$R_\odot$. The CME kinematics can be divided into three phases: initial acceleration, main acceleration, and propagation with constant velocity.  After the CME onset GOES registered a sub-A-class flare.
\end{abstract}


\section{Introduction}
Coronal Mass Ejections (CME) are ejections of  coronal plasma into the interplanetary space. CME mass lies in the interval $10^{12}$~--~$10^{16}$~g \citep{Vourlidas2010}, and CME size could be larger than solar radii ($R_\odot$). CMEs occur due to the large-scale processes of energy release, and that is why CME investigations are important for solar physics. CME investigations are also critically important for solar-terrestrial issues because CMEs, which reach the Earth, influence Earth's magnetosphere and satellite operation \citep{Schwenn2006, Pulkkinen2007}.

Approximately 30\% of  CMEs have a three-part structure: bright core, dark cavity, and bright frontal loop \citep{Illing1985, Webb1987}. Perhaps more CMEs have such structure, but due to projection effects, it is unobservable. It is considered that the bright core is the prominence, which erupted with the CME \citep{House1981}; the dark cavity is the less dense plasma surrounding the prominence; and the bright frontal loop is the coronal plasma, which is amassed during the CME motion \citep{Forbes2000}.  

Primary CME acceleration occurs at distances of 1.4--4.5~$R_{\odot}$ \citep{Zhang2001, Zhang2004, Gallagher2003}. At distances higher than 5~$R_\odot$, CMEs can accelerate, decelerate, or move with constant velocity. In general, CMEs with velocity greater than 400~km~s$^{-1}$ decelerate, and with less than 400~km~s$^{-1}$ accelerate. It is considered that fast CMEs move faster than solar wind, and the wind decelerates them; and slow CMEs move slower than solar wind, and the wind accelerates them \citep{Yashiro2004}. CME velocity lies in the range 20--3500~km~s$^{-1}$ \citep{Gopalswamy2010}. Non-impulsive CME acceleration lies in the range -60--+40~m~s$^2$ \citep{Yashiro2004}, and impulsive acceleration is of an order of 500~m~s$^{-2}$ \citep{Zhang2001}.  

CME behavior depends on the phase of the solar cycle. In the solar maximum CMEs occur more often than in the minimum \citep{Robbrecht2009}. In the maximum, average CME velocity is about 500~km~s$^{-1}$ and in the minimum 250~km~s$^{-1}$. Moreover, in the maximum, CMEs are observed  at all latitudes, and in the minimum, they are close to the ecliptic plane \citep{Yashiro2004}.

There is no clear connection between CMEs and other solar phenomena. Some CMEs are associated with flares, some with erupting prominences, some with both flares and erupting prominences \citep{Webb2012}, and some CMEs, known as ``stealth CMEs'', are not associated with any visible changes on the solar surface \citep{Ma2010}. 

For the first time, CMEs were observed with white-light coronagraphs on board the satellite \textit{OSO-7} in 1971 \citep{Tousey1973}. Early CME investigations were carried out with white-light coronagraphs on board \textit{Skylab} (1973--1974, \citet{Munro1979}), \textit{P78-1/Solwind} (1979-1985, \citet{Sheeley1980}), and \textit{Solar Maximum Mission} (1980; 1984--1989; \citet{Hundhausen1999}). 

The modern era of CME investigations started with the launch of the \textit{SOHO} satellite (1995; \citet{Domingo1995}). SOHO carries white-light coronagraphs LASCO/C1, C2, C3, which take images of the solar corona at distances 1.1--30~$R_{\odot}$ (C1: 1.1--3~$R_{\odot}$; C2: 2--6~$R_{\odot}$; C3: 4--30~$R_{\odot}$). Coronagraph C1 operated from 1995 till 1998, and coronagraphs C2 and C3 still work. LASCO coronagraphs have obtained large volumes of data, and most of the knowledge about CME behavior we have thanks to these instruments.

The next  step in experimental CME investigation was the launch of  \textit{STEREO} satellites (2006; \citet{Howard2008}). \textit{STEREO} satellites move along the Earth's orbit: one  ahead of the Earth (\textit{STEREO-A}), and the other in the opposite direction (\textit{STEREO-B}). Both \textit{STEREO} satellites carry white-light coronagraps: COR1 and COR2. COR1 builds solar corona images at distances 1.3~--~4~$R_\odot$ and COR2 at distances 2~--~15~$R_\odot$. LASCO and \textit{STEREO} coronagraphs observe the Sun from three different viewpoints, allowing reconstruction of three-dimensional CME trajectory and structure. Furthermore, \textit{STEREO} coronagraphs can see CMEs, that are not seen with LASCO. More details about CME observations can be read in the review \citet{Webb2012}.

Today, the CME behavior is studied best at distances greater than 2~$R_\odot$ from the Sun's center. There are a small number of observations in the distance range 1.2--2~$R_\odot$, and this distance range is studied less. However, this distance range is the most interesting because it is the range where CME forms, accelerates, and evolves.

In this work, we present results of the observations of a CME, which occurred on May 13, 2009. The most important feature of these observations is that the CME was observed from the very early stage (the solar surface) up to the distance of 15~$R_\odot$. Below 2~$R_\odot$, we used the data from TESIS EUV telescopes obtained in Fe 171~\AA\ and He 304~\AA\ lines \citep{Kuzin2011}, and above 2~$R_\odot$, we used the data from LASCO C2 and C3 coronagraphs. The aim of this work is to describe the formation, early evolution, and propagation of this CME.

\section{Experimental Data}

TESIS is an instrument assembly employed for solar corona investigations in the extreme ultra-violet (EUV) and soft X-ray wavelength ranges. TESIS was launched on board the \textit{CORONAS-PHOTON} satellite in 2009  \citep{Kotov2011}. TESIS included EUV telescopes that built images in Fe~171~\AA\ and He~304~\AA\ lines. The telescopes had $1^{\circ}$ field of view and $1.7^{\prime\prime}$ angular resolution. In the synoptic observation mode, the telescope's cadence  was 10--15 minutes.

The main difficulties in the construction of far corona EUV images are small field of view and limited dynamic range of the telescopes. The field of view of TESIS EUV telescopes reached distances of 2~$R_\odot$ from the Sun's center, but their dynamic range was not high enough to build an image of the far corona.  On images with long exposure, the far corona is visible, but the solar disk is overexposed. Conversely, on images with short exposure, the solar disk is not overexposed, but the far corona is too faint. 

\begin{figure*}[t]
\centering
\includegraphics[width = 0.7\textwidth]{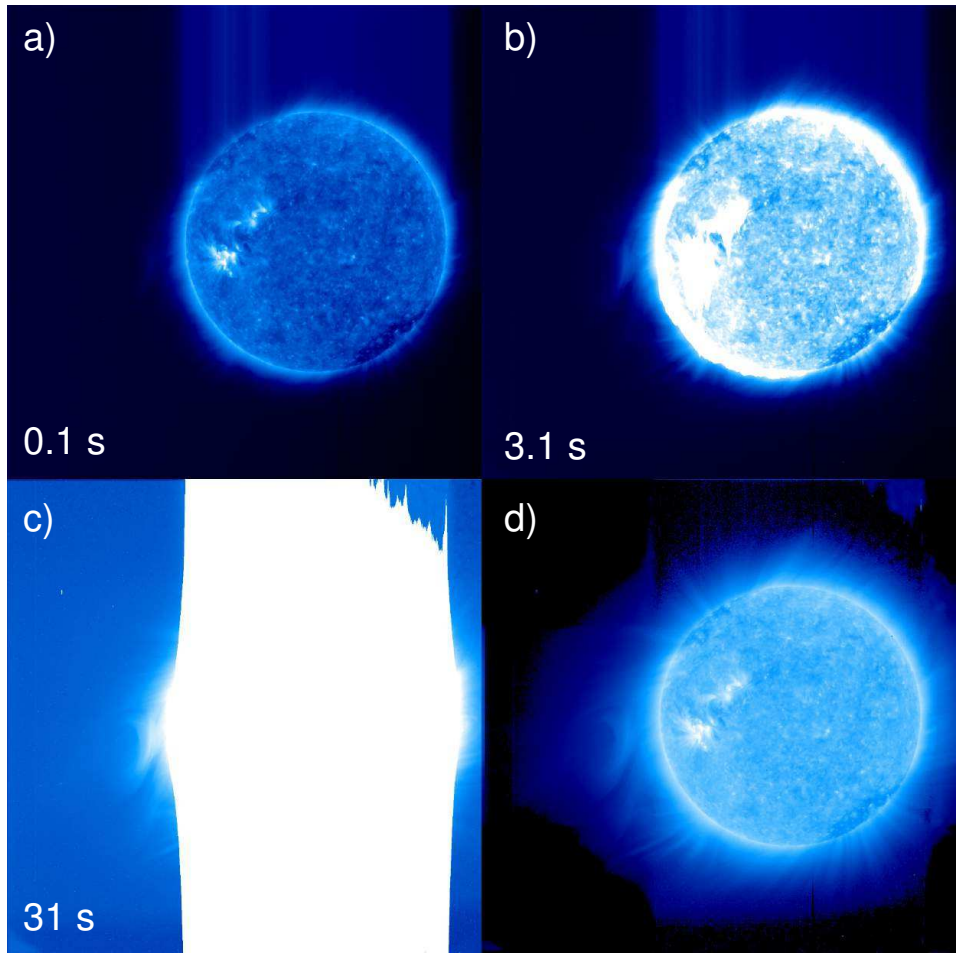}
\caption{Processing of the TESIS 171~\AA\ images. a, b, c --- raw TESIS images with different exposure times (a---0.1~s, b---3.1~s, c---31~s). d --- Composite image. We used logarithmic scale for all four images in this figure.}
\label{F:prep}
\end{figure*}

In order to see the far corona with the TESIS Fe~171~\AA\ telescope, we developed special observational program. During this program, the telescope took images with different exposure times: short (0.1 s), medium (3.1 s), and long (31 s) (see Figure~\ref{F:prep}a,b,c). As you can see a lot of the pixels on the 31~s image are saturated. The idea of our method is to substitute saturated pixels on the 31~s image with non-saturated pixels from the 0.1~s or 3.1~s images. We do this with the following algorithm:
\begin{enumerate}
\item For each image we performed an initial processing: removing of hot pixels, background, flat-field adjustment, etc. For more details see \citet{Kuzin2011b}. At this step we do not remove ``the trace'' --- blue stripes to the north from the Sun (see Figure~\ref{F:prep}a,b).
\item We co-aligned the images using correlation techniques. 
\item We scaled the  intensities of the 0.1~s and 3.1~s  images to the intensity of the 31~s image. The scaling coefficient is the same for all pixels on the same image, but different for the 0.1~s and 3.1~s images.
\item We made a composite image out of these three images. If pixel is not saturated on the 31~s image, we pick its intensity as the composite image intensity. If pixel is saturated on the 31~s image, but not saturated on the 3.1~s image, we pick 3.1~s image intensity. In other cases we pick 0.1~s image intensity (see Table~\ref{T:Prep}).
\item We removed the trace. During the read-out the CCD-matrix was exposed to the Sun, and this additional exposure generated trace. The trace is a linear function of the signal from the Sun, and we can calculate it analytically. The trace removal algorithm is described in \citet{Kuzin2011b}.
\end{enumerate}
The result is shown in Figure~\ref{F:prep}d. We must stress that this method doesn't recover the far corona information for all pixels on the composite image. The pixels on the 31~s image to the north and the south form the Sun are saturated, and we used the intensity from the 3.1~s image for these regions. As a consequence, the noise dominates over signal at closer distances from the Sun in the north and the south regions than in the west and east ones. We were lucky that the CME, analyzed in this work, occurred at the eastern edge of the image.
\begin{table*}
\centering
\caption{Formation of the composite image.}
\begin{tabular}{cccc}
\tableline 
Pixel on 0.1 s image & Pixel on 3.1 s image & Pixel on 31 s image & Composite image pixel\\ 
\tableline 
Saturated     & Saturated     & Saturated     & 0.1 s image \\ 
Not saturated & Saturated     & Saturated     & 0.1 s image \\ 
Not saturated & Not saturated & Saturated     & 3.1 s image \\ 
Not saturated & Not saturated & Not saturated & 31  s image \\ 
\tableline 
\end{tabular} 
\label{T:Prep}
\end{table*}

\begin{figure*}[hbt]
\centering
\includegraphics[width = 0.7\textwidth]{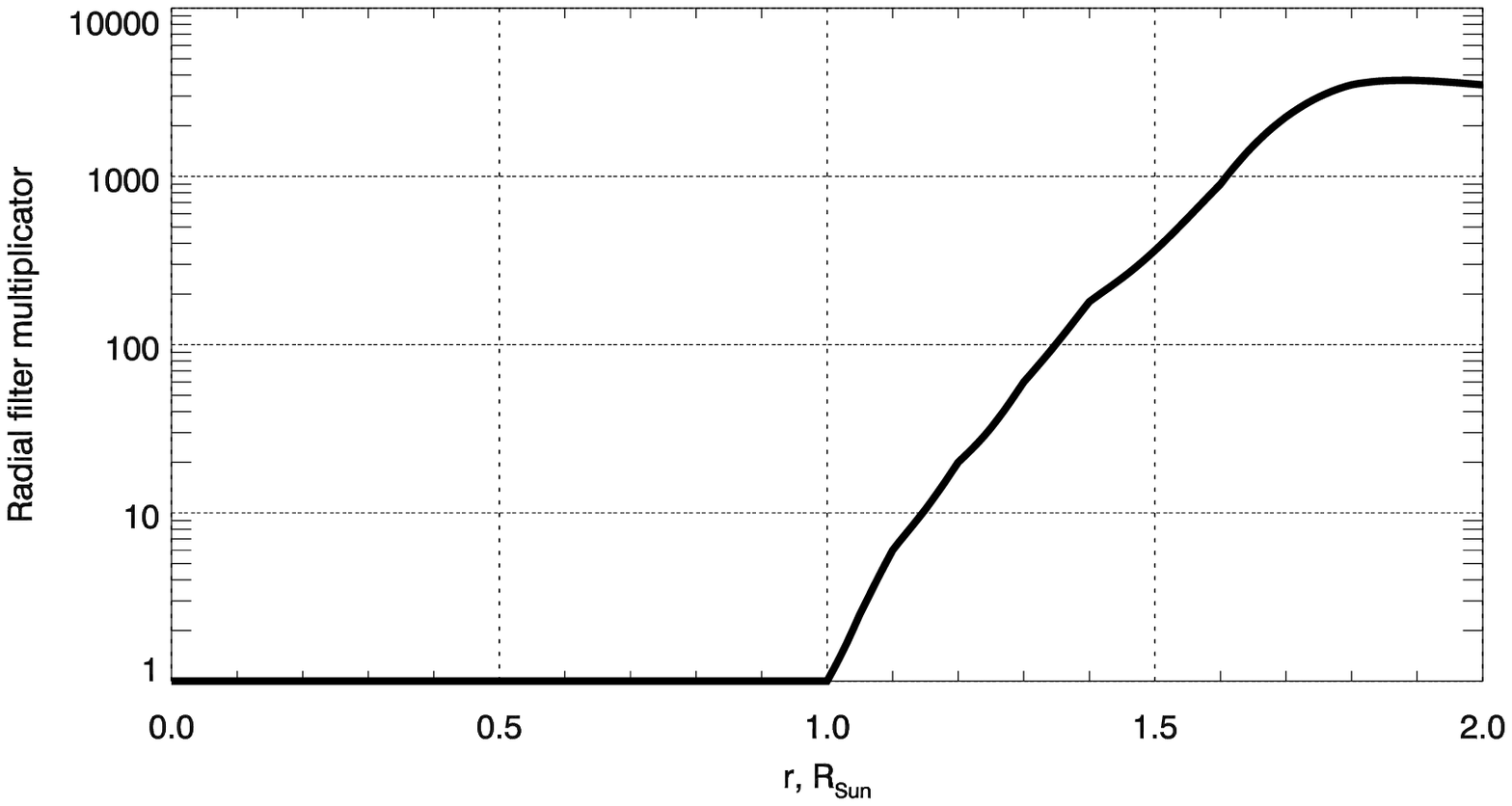}
\caption{Radial filter function.}
\label{F:radial_filter}
\end{figure*}

\begin{figure*}[hbt]
\centering
\includegraphics[width = 0.9\textwidth]{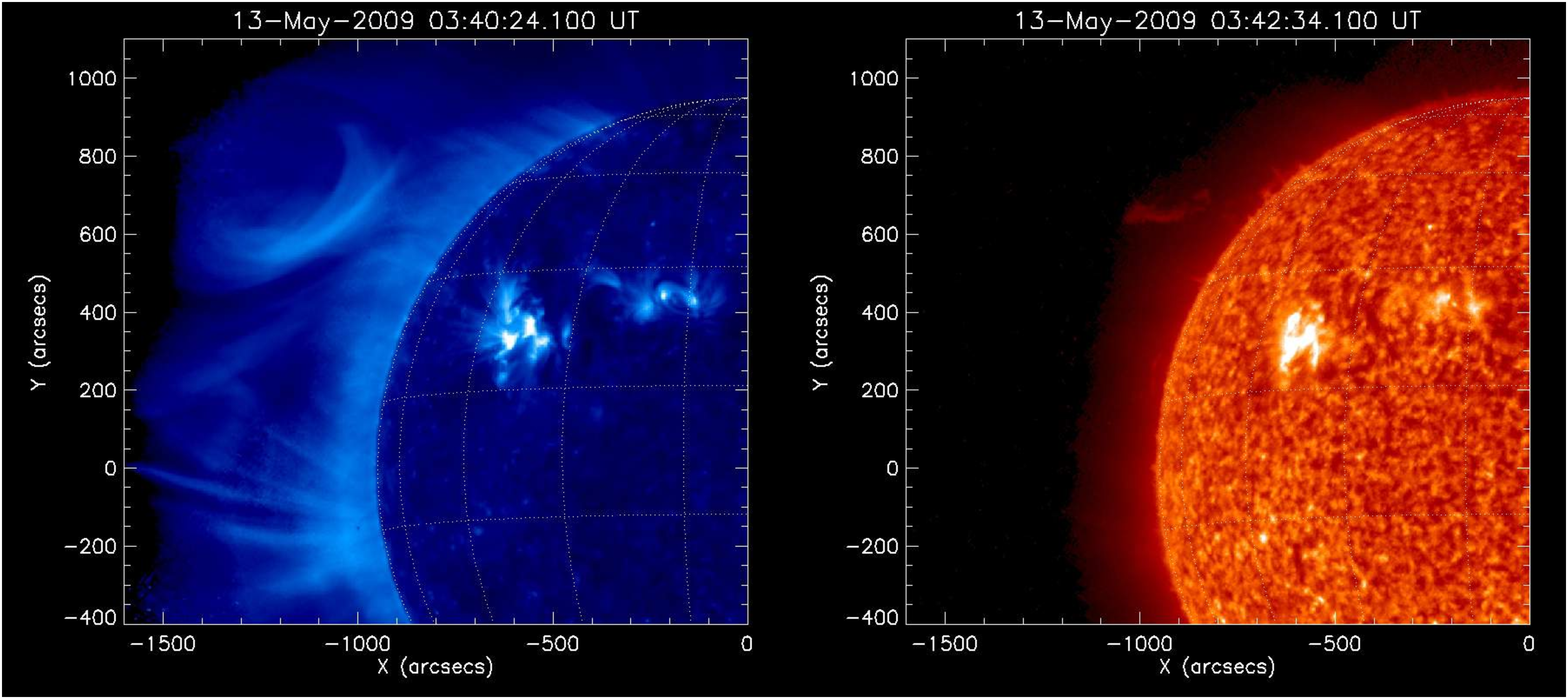}
\caption{CME in the low corona, which occurred on May 13, 2009. Left --- TESIS Fe~171~\AA\ image with applied radial filter; right --- TESIS He~304~\AA\ image.}
\label{F:radial_filter_image}
\end{figure*}

The intensity of the far corona is significantly lower than the solar disk intensity. In order to show all layers of the solar corona on one image, we need to artificially increase the intensity of far corona. In this work, we have applied a radial filter for Fe~171~\AA\ images --- we have multiplied the signal by a function, that depends on the distance from the Sun's center.  Inside the solar disk, this function  equals unity. Outside the solar disk, we picked the function values to make the far corona intensity to be the same order of magnitude as the solar disk intensity. The radial filter function is shown in Figure~\ref{F:radial_filter}. The image after applying the radial filter is shown in Figure~\ref{F:radial_filter_image}.

In this work, we investigate a CME, which occurred on May 13, 2009 at 00:10~UT close to the solar limb. The CME was observed with the TESIS Fe~171~\AA\ telescope in the far corona observation mode. Below 2~$R_\odot$, we used the data from TESIS EUV telescopes obtained in the Fe 171~\AA\ and He 304~\AA\ lines, and above 2~$R_\odot$, we used the data from LASCO C2 and C3 coronagraphs. We made a movie of these observations, and we strongly recommend the reader to watch it before proceeding further (see Figure~\ref{F:Fe_C2}).

STEREO coronographs also observed this CME. The main focus of our work is EUV observations of the CME in the low corona. LASCO better complements TESIS observations than STEREO, because LASCO has the same viewpoint as TESIS. That is why we do not use STEREO data in our research.

\section{Results}
\subsection{General Properties}

The observed CME is relatively slow --- it takes it 15~hours to travel from the solar surface to the LASCO/C2 field of view. Its angular size is 15$^\circ$. The CME started at 50$^\circ$ latitude and then moved along a curved trajectory to the ecliptic plane.

\begin{figure*}[hbt]
\centering
\includegraphics[width = 0.95\textwidth]{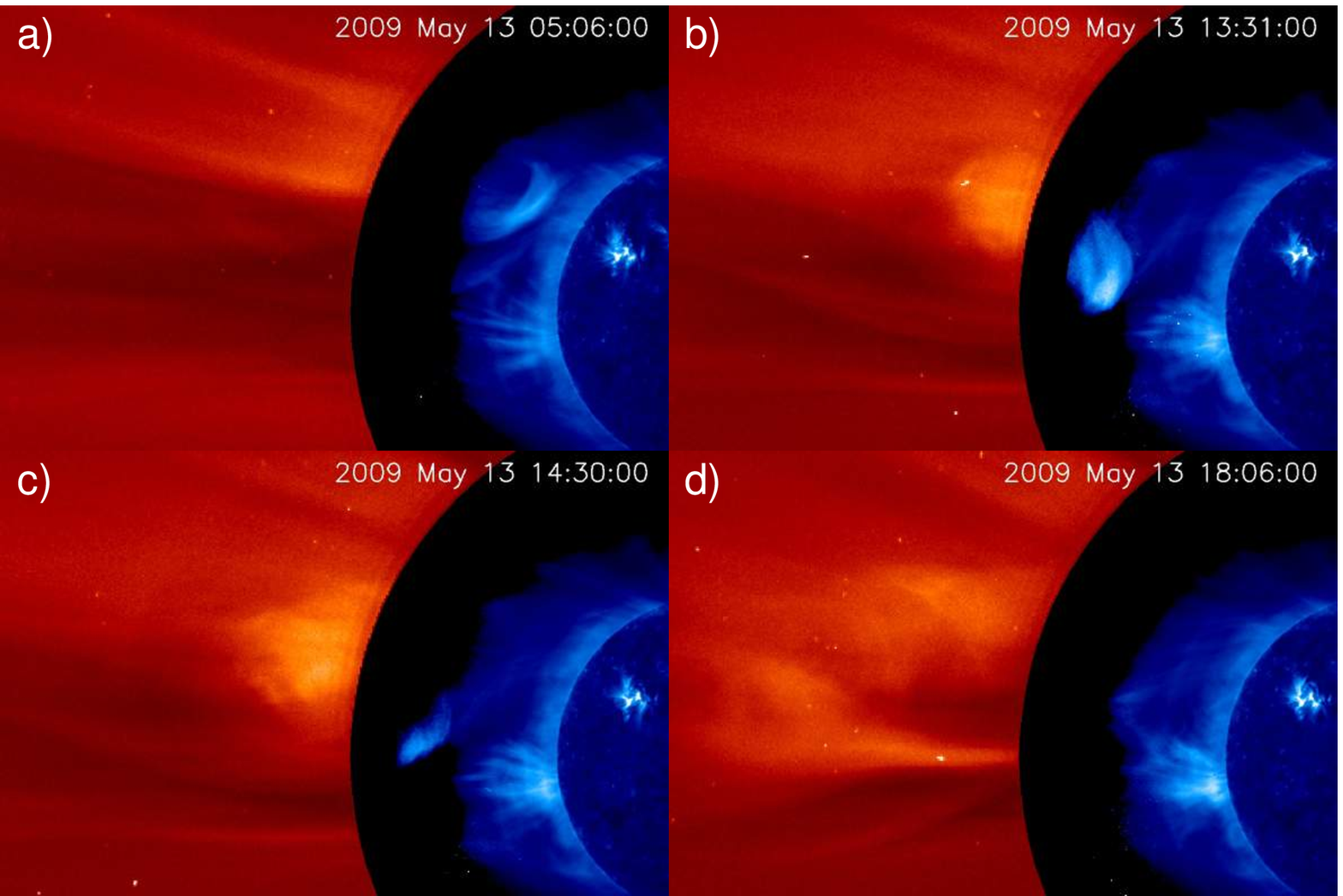}
\caption{Composite LASCO C2 (red) and TESIS Fe~171~\AA\ (blue) images.}
\label{F:Fe_C2}
\end{figure*}

The observed CME does not have the common three-part structure (bright core, dark cavity, bright frontal loop). In all stages of its evolution, the CME had a U-shaped structure, which formed at the height of 0.2--0.5~$R_\odot$ from the solar surface. The CME's U-shaped structure extended to the LASCO field of view (see~Figure~\ref{F:Fe_C2}a), which means that the CME had an open magnetic field structure.

In the early stages of the evolution, the CME was located at high latitudes, and its `leg'---the magnetic field line that connects the CME to the Sun---connected with the solar surface at high latitudes (see~Figure~\ref{F:Fe_C2}a). In the late stages of the evolution, the CME reached the ecliptic plane, and its `leg' connected with the solar surface at the equator (see~Figure~\ref{F:Fe_C2}b,c). Somewhere at distances 1.8~--~2.3~$R_{\odot}$ from the Sun's center the CME changed its  `leg'. Unfortunately, at this distance range, TESIS images are too blurry, and we can not distinguish this process.

When the CME passes from TESIS Fe 171~\AA\ to LASCO C2 field of view, we see complete correspondence between the bright structures on the TESIS and LASCO images (see~Figure~\ref{F:Fe_C2}b,c). Since bright structures on LASCO images are density structure, then the CME on TESIS Fe 171~\AA\ images is also a density structure.

\begin{figure*}[hbt]
\centering
\includegraphics[width = 0.9\textwidth]{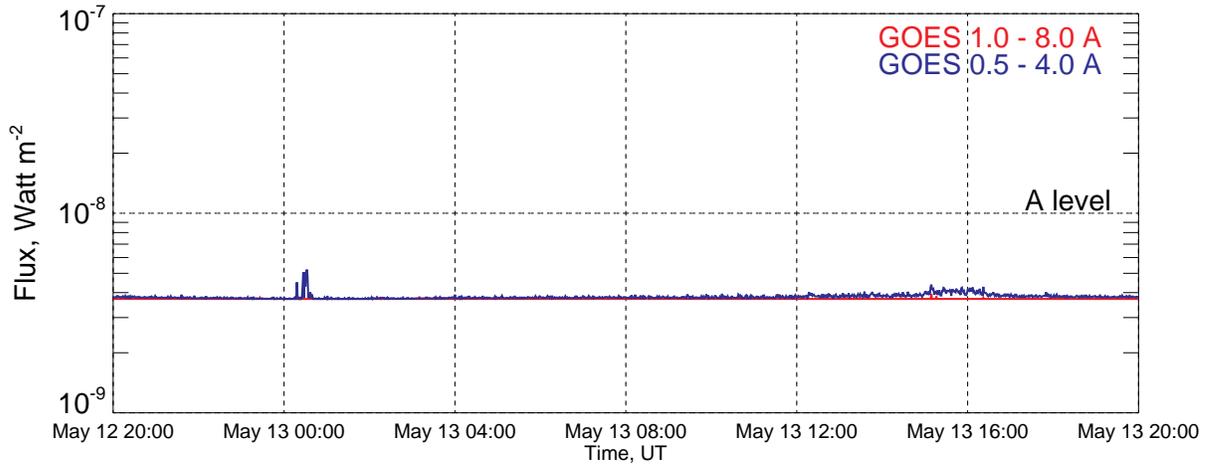}
\caption{GOES flux. Red: the GOES 1.0--8.0~\AA\ channel; blue: the GOES 0.5--4.0~\AA\ channel.}
\label{F:GOES}
\end{figure*}

After the CME started, a flare occurred at 00:30 UT (see~Figure~\ref{F:GOES}). GOES flux in the 0.5--4~\AA\ channel did not exceed A-level, the flux in the 1--8~\AA\ channel was below the sensitivity threshold. We cannot prove that this flare is related to the CME, so the flare could occur in some other remote place on the Sun.

\subsection{Prominence}

\begin{figure*}[hbt]
\centering
\includegraphics[width = 0.9\textwidth]{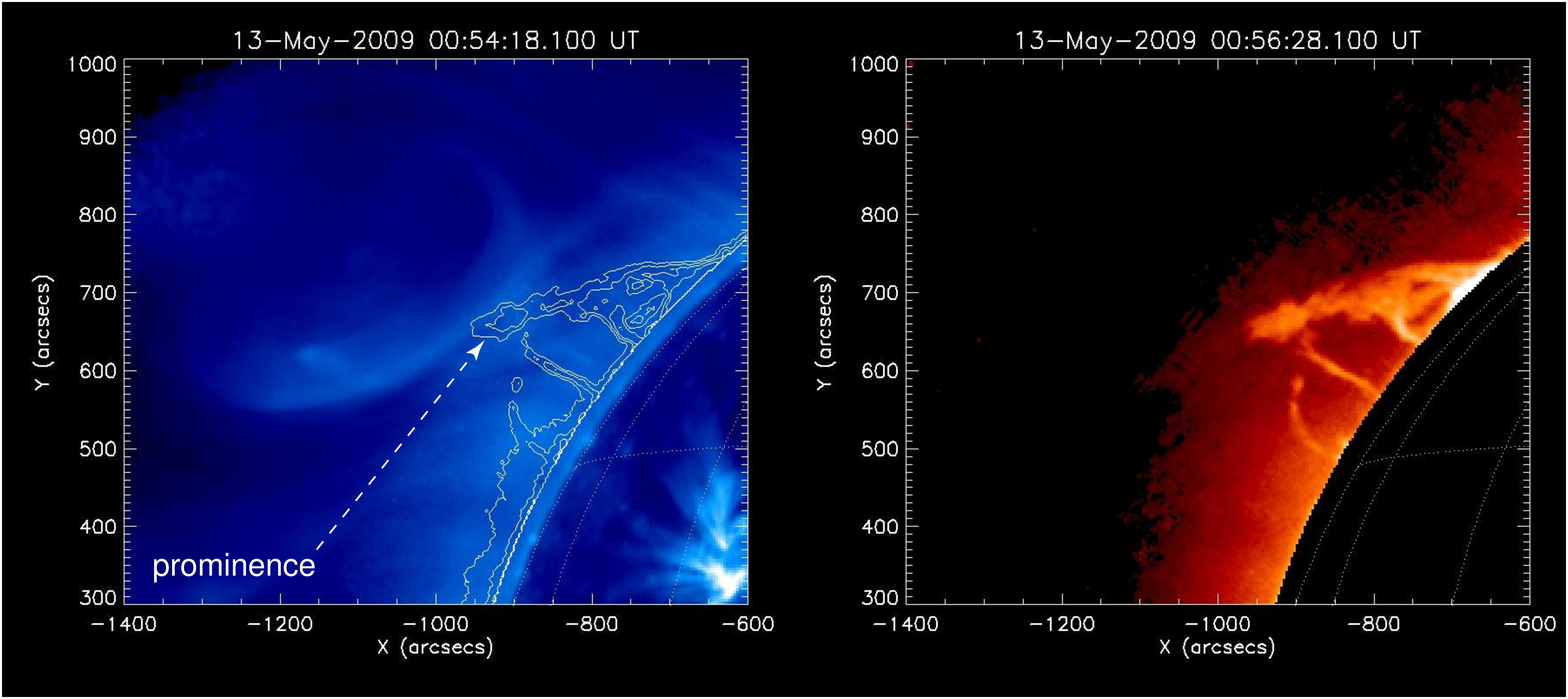}
\caption{Prominence location relative to U-shape structure. Left: TESIS Fe~171~\AA\ images with applied radial filter. The prominence is overlayed with contours. Right: TESIS He~304~\AA\ images. We put an artificial occulting disk on He~304~\AA\ image in order to improve the prominence visibility.}
\label{F:prominence_below_the_cavity}
\end{figure*}

\begin{figure*}[hbt]
\centering
\includegraphics[width = 0.95\textwidth]{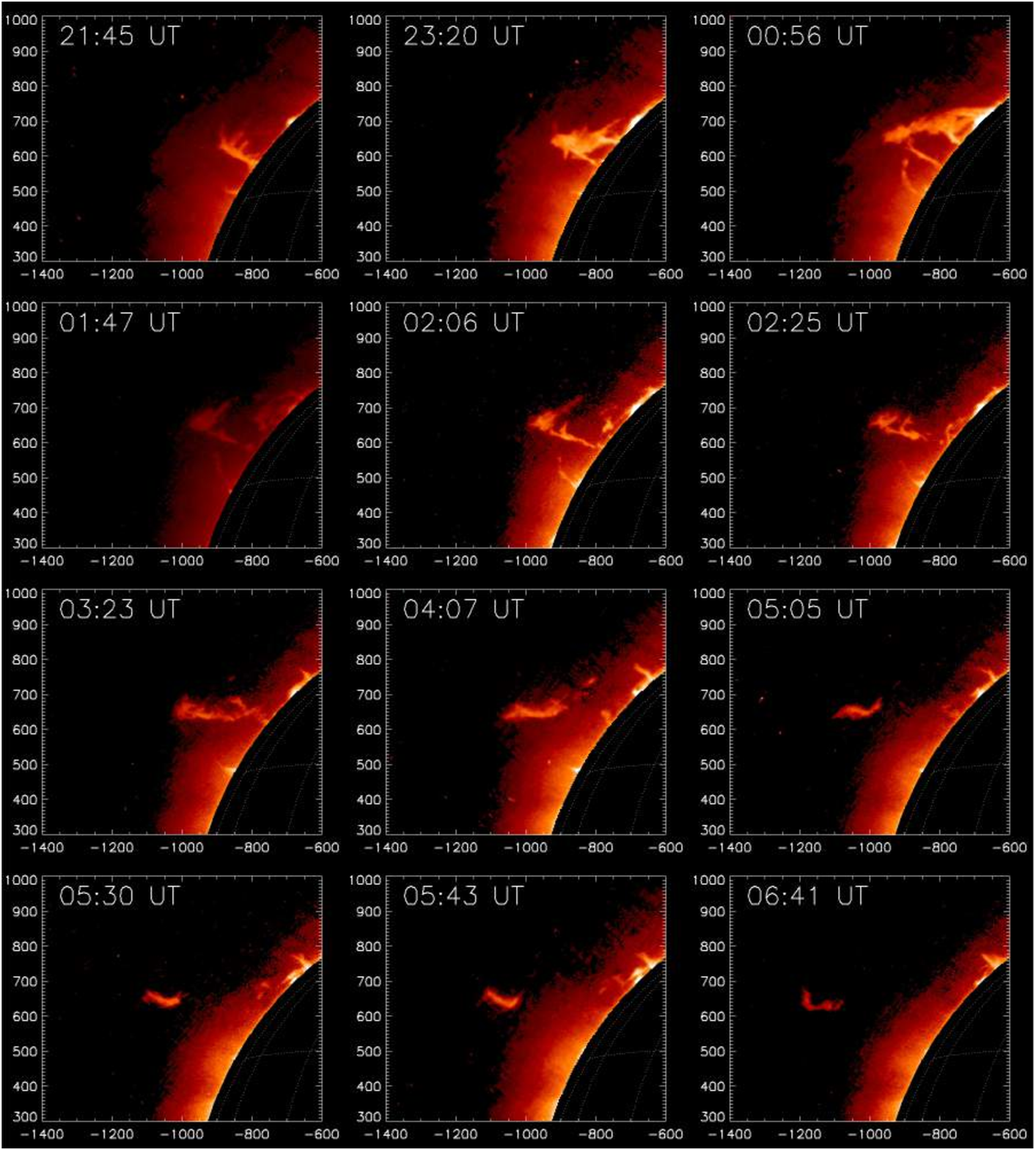}
\caption{Prominence evolution. The coordinates are measured in arc seconds. We put an artificial occulting disk on the He~304~\AA\ images in order to improve the prominence visibility.}
\label{F:prominence_life}
\end{figure*}

The CME was associated with an erupting prominence, which was observed in TESIS He~304~\AA\ images. The prominence was below the U-shaped structure, close to the magnetic X-point (see~Figure~\ref{F:prominence_below_the_cavity}). Just before the eruption, the prominence broke into two parts: one fell on the Sun; the other left the Sun with the CME (see~Figure~\ref{F:prominence_life}). The prominence was observed up to the height of 0.4~$R_\odot$ above the solar surface, where it faded away.

\subsection{CME mass}

\begin{figure*}[hbt]
\centering
\includegraphics[width = 0.9\textwidth]{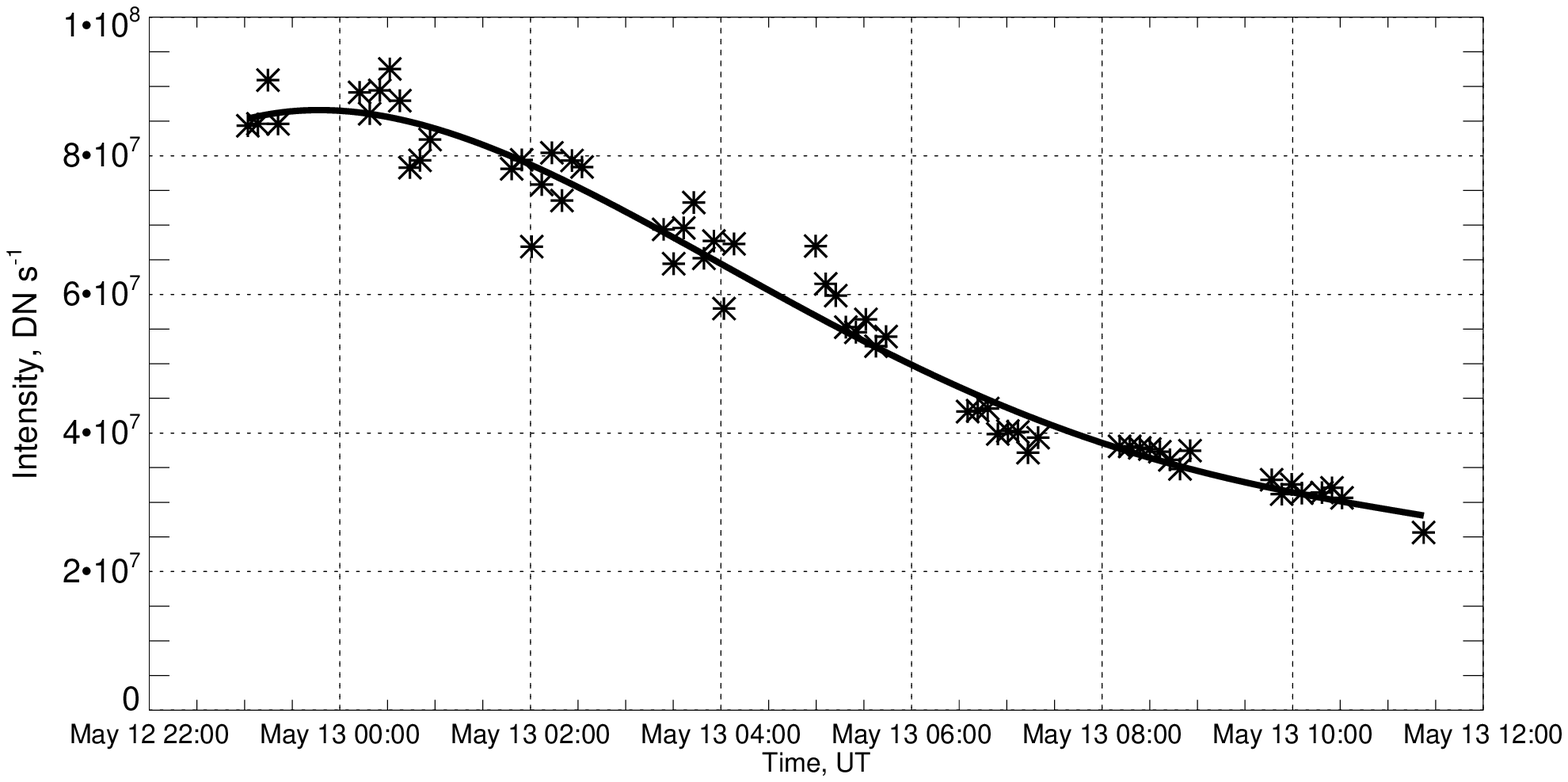}
\includegraphics[width = 0.9\textwidth]{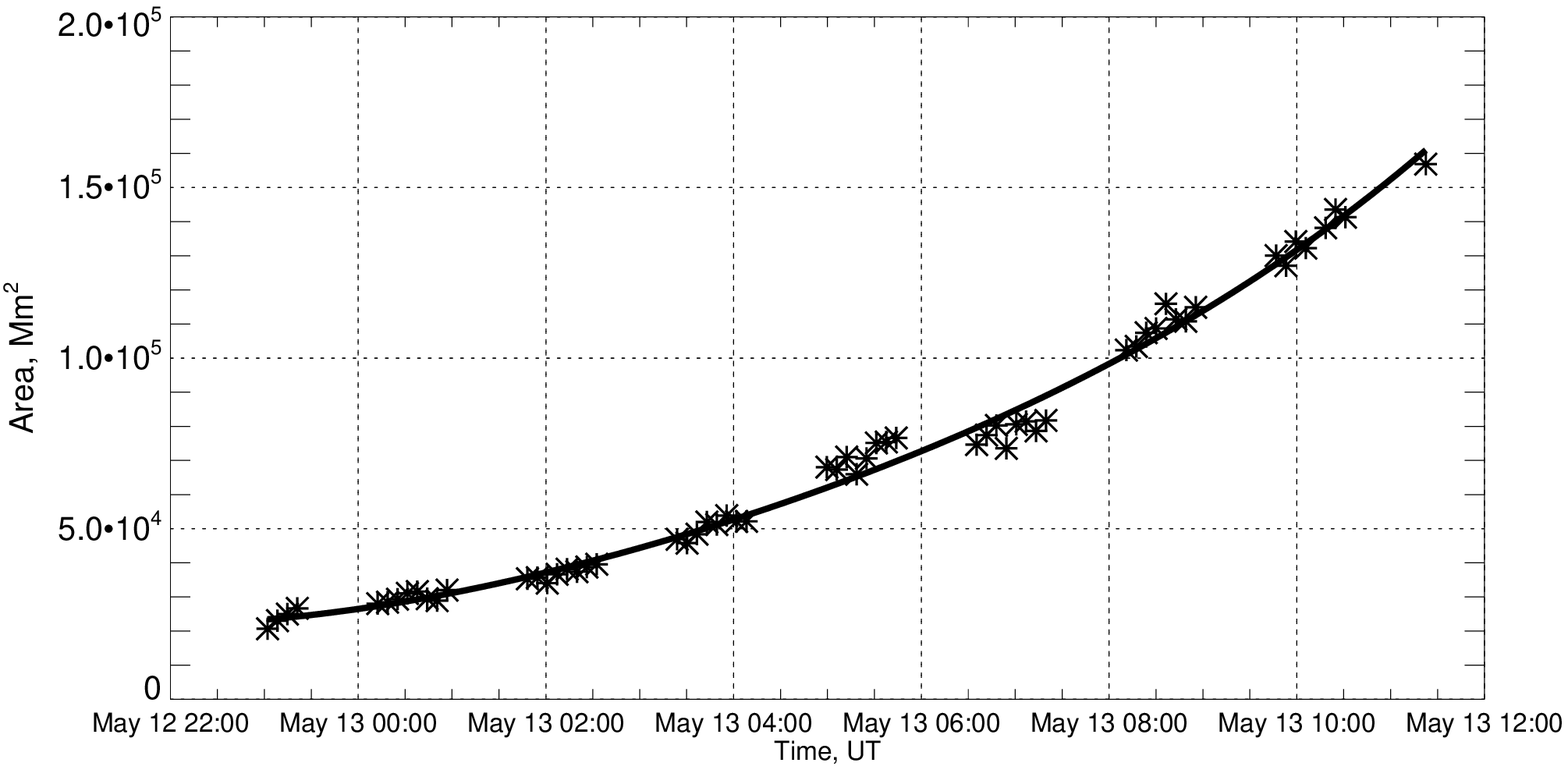}
\caption{Top panel: the CME intensity on TESIS Fe~171~\AA\ images. Bottom panel: CME area on TESIS Fe~171~\AA\ images.}
\label{F:CME_flux}
\end{figure*}

In order to estimate the CME mass, we measured CME intensity and area on TESIS Fe~171~\AA\ images. The intensity ($I$) decreased by 3.6 times, and the area ($A$) increased by 6.5 times (see Figure~\ref{F:CME_flux}). 

The CME intensity on the TESIS Fe~171~\AA\ images equals
\begin{equation}
I = \int G(T) DEM(T)\ dT,
\end{equation}
where $G(T)$---the temperature response of the TESIS Fe~171~\AA\ telescope, and $DEM(T)$---the CME differential emission measure.

For an estimation, we assume that the CME was isothermal and its temperature was constant. In this case,
\begin{equation}
I = G(T_0) EM \sim n_e^2 V = (\frac{N}{V})^2 V = \frac{N^2}{V},
\end{equation}
where $T_0$---the CME temperature; $EM$---the CME emission measure; $n_e$---the CME electron density; $V$---the CME volume; $N$---the number of electrons in the CME.  The CME mass $M$ is proportional to the number of electrons $N$:
\begin{equation}
M \sim N \sim \sqrt{I V}
\end{equation}

Let us consider two cases of the relationship between CME volume and area:
\begin{enumerate}
\item $V \sim A$ --- the CME did not expand in the direction perpendicular to the image plane;
\item $V \sim A^{3/2}$ --- the CME expanded evenly in all directions.
\end{enumerate}

In the first (two-dimensional) case:
\begin{equation}
M_{2D} \sim \sqrt{I V} \sim \sqrt{I A}
\end{equation}

In the second (three-dimensional) case:
\begin{equation}
M_{3D} \sim \sqrt{I V} \sim \sqrt{I A^{3/2}}
\end{equation}

\begin{figure*}[hbt]
\centering
\includegraphics[width = 0.9\textwidth]{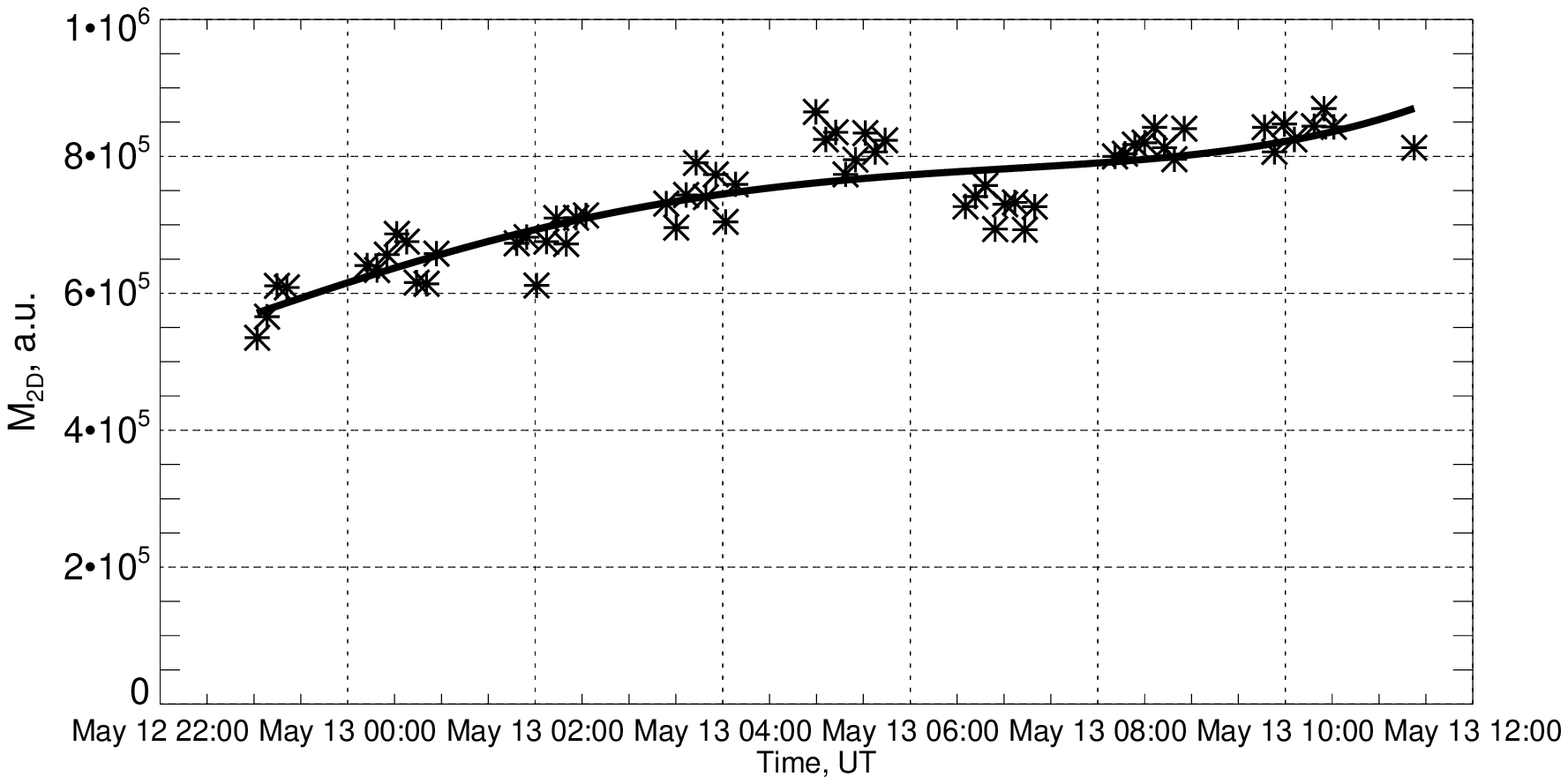}
\includegraphics[width = 0.9\textwidth]{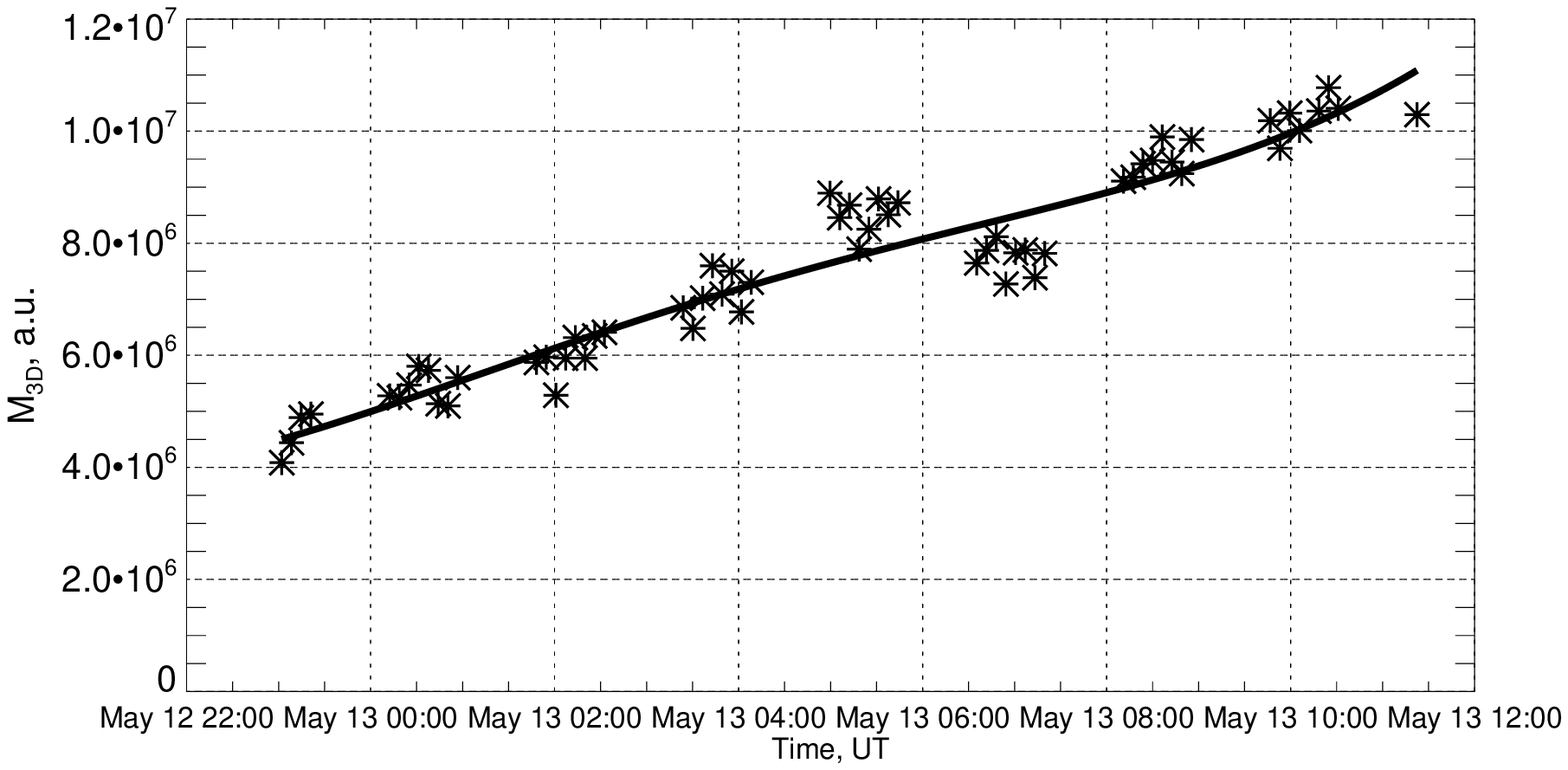}
\caption{Top panel: the CME mass in two-dimensional case; bottom panel: the CME mass in three-dimensional case.}
\label{F:CME_mass_self}
\end{figure*}

For each of these cases, we calculated the CME mass using measured values of $I$ and $A$. In the two-dimensional case the CME mass increased by 30~\%, in the three-dimensional case by 150~\% (see~Figure~\ref{F:CME_mass_self}). In reality, we deal with an intermediate case, and the real CME mass dynamics is somewhere between these two cases. Since in both cases the mass increased, the real CME mass should increased by 30--150~\%.

\subsection{Kinematics}

\begin{figure*}[!t]
\centering
\includegraphics[width = 0.9\textwidth]{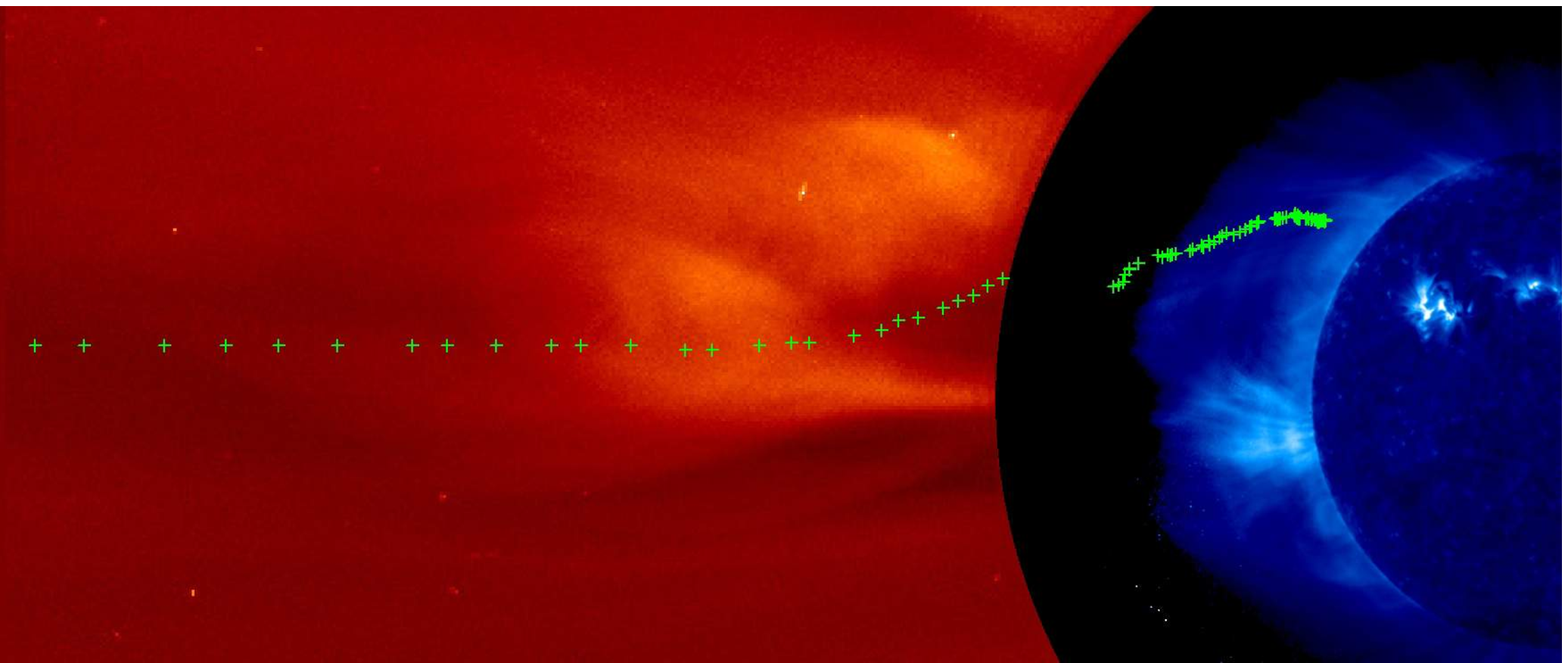}
\caption{CME trajectory. Left: LASCO C2 image; right: TESIS Fe~171~\AA\ image. Green crosses designate CME's position at different times.}
\label{F:Trajectory}
\end{figure*}

We measured the coordinates of the CME's frontal bright edge on TESIS and LASCO images. For LASCO C3 images, we measured coordinates until the CME could not be distinguished from the background. The CME had curved trajectory---its helio-latitude decreased with time (see Figure~\ref{F:Trajectory}). The CME originated at 50$^\circ$ latitude  and reached ecliptic plane at 2.5~$R_\odot$. 

\begin{figure*}[hbt]
\centering
\includegraphics[width = 0.9\textwidth]{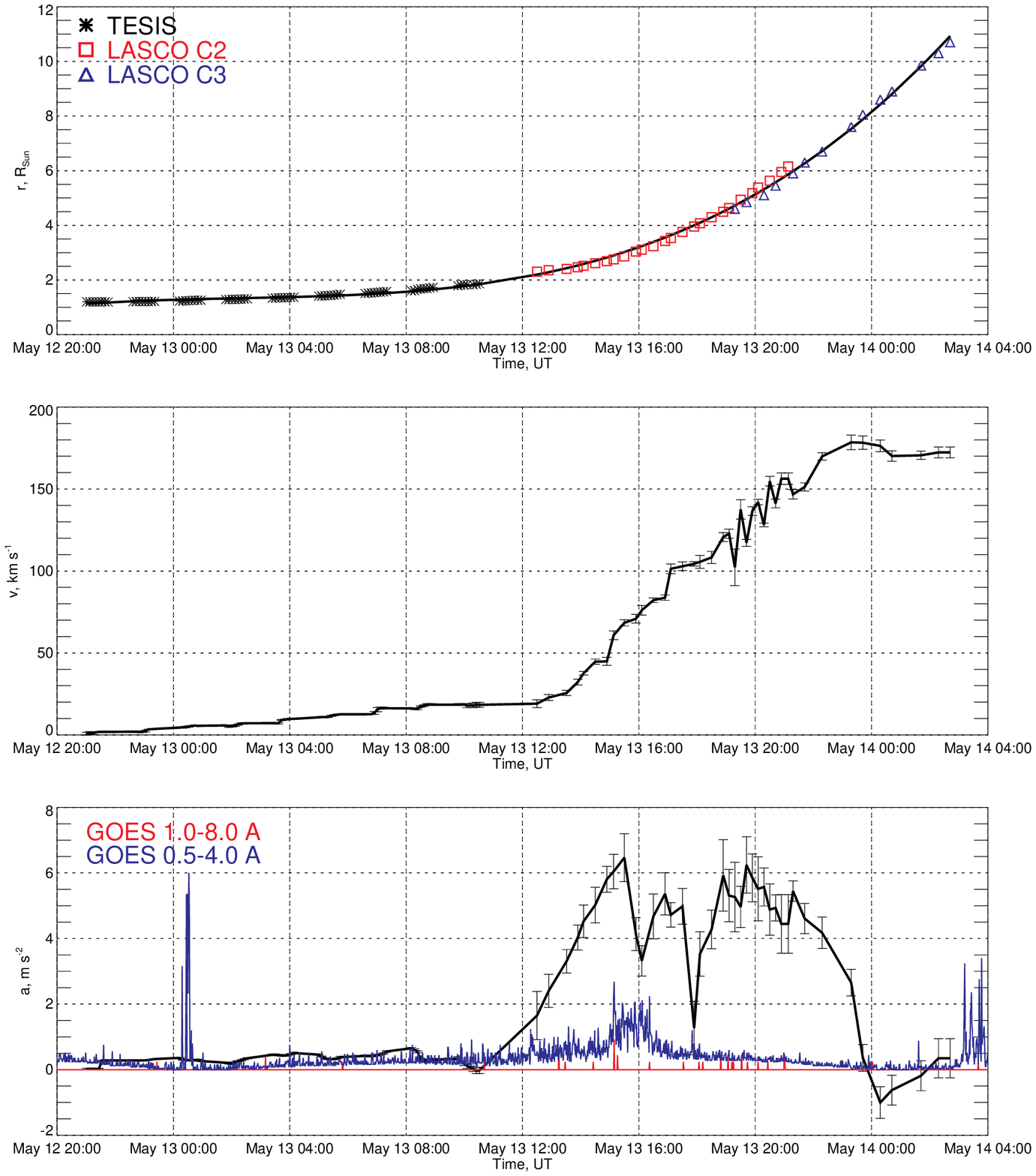}
\caption{CME kinematics: $r$: distance to the Sun's center; $v$: radial CME velocity, $a$: radial CME acceleration; $R_{Sun}$: solar radius. On the acceleration plot (bottom panel), GOES flux is over plotted.}
\label{F:Kinematics}
\end{figure*}

Figure~\ref{F:Kinematics} shows the dependencies of the CME's distance from the Sun's center, velocity, and acceleration on time. The CME kinematics could be divided into three phases: initial acceleration, main acceleration, and steady movement.

During the first phase, the CME had an acceleration 0.5~m~s$^{-2}$, velocity 20~km~s$^{-1}$, and reached the height of 1~$R_\odot$ above the solar surface within 14~hours. During the second phase, the CME acceleration increased to 5~m~s$^{-2}$, and the velocity increased from 20 to 170~~km~s$^{-1}$. The second phase took place at distances 2--6~$R_\odot$ over 12~hours. During the third phase, the CME moved with a constant velocity of 170~~km~s$^{-1}$. 

\section{Discussion}

\subsection{Prominence}

\begin{figure*}[hbt]
\centering
\includegraphics[width = 0.6\textwidth]{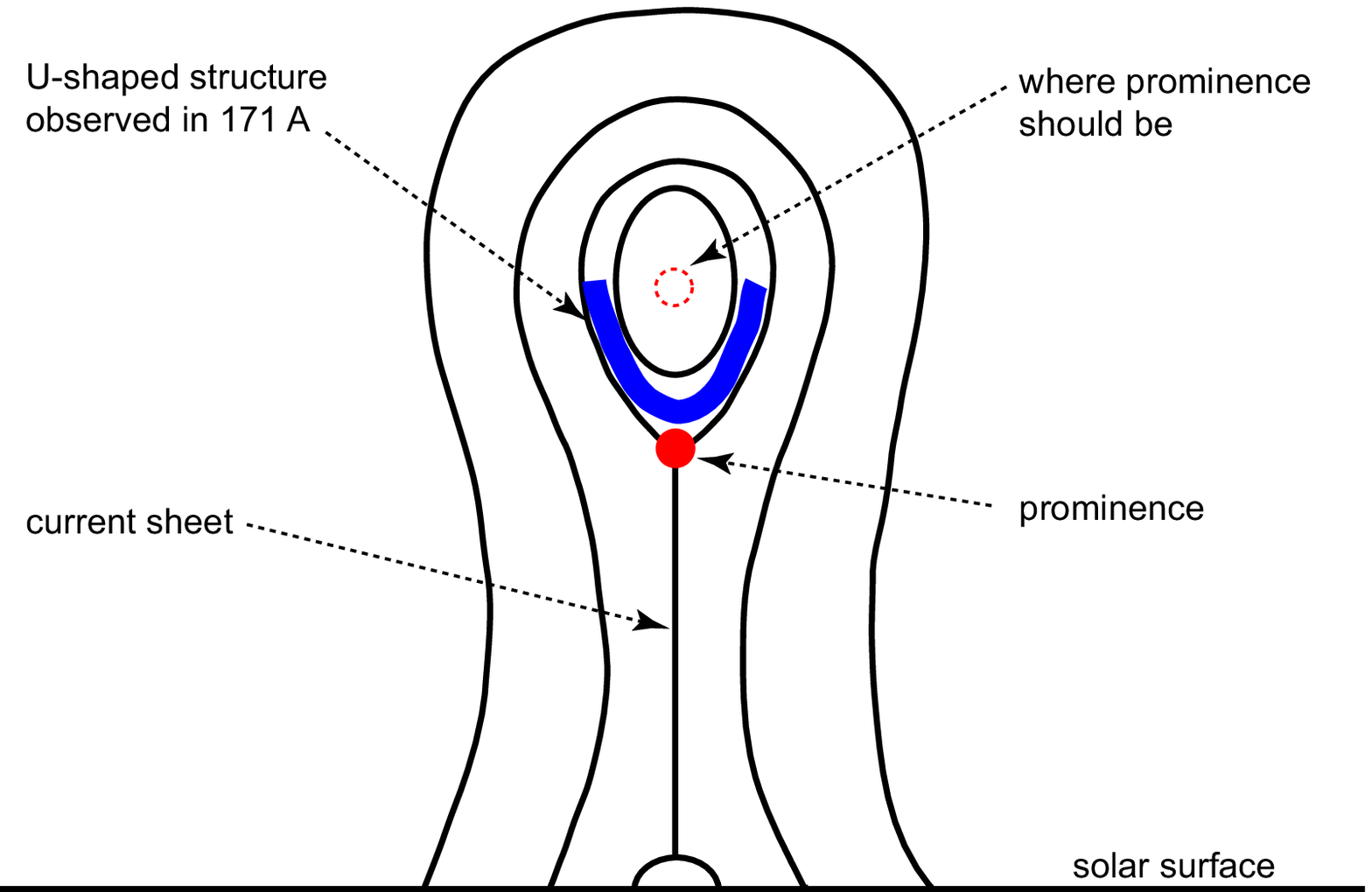}
\caption{CME standard model and the real prominence location.}
\label{F:standard_model}
\end{figure*}

According to the standard CME model, the CME has structure, which is shown in Figure~\ref{F:standard_model} \citep{Carmichael1964, Sturrock1966, Hirayama1974, Kopp1976}. The prominence should be located above the current sheet and be surrounded by circular magnetic field lines. Under the prominence in the current sheet the reconnection happens. Plasma outflow from the reconnection region pushes the prominence up, and the CME erupts. This outflow also fills magnetic field lines under the prominence and forms the observed in the 171~\AA\ line U-shaped structure. So, in the standard CME model, the prominence is located above the U-shaped structure. This contradicts the presented observations, in which the prominence is located in the lowest part of the U-shaped structure close to the magnetic X-point. 

\begin{figure*}[hbt]
\centering
\includegraphics[width = 0.8\textwidth]{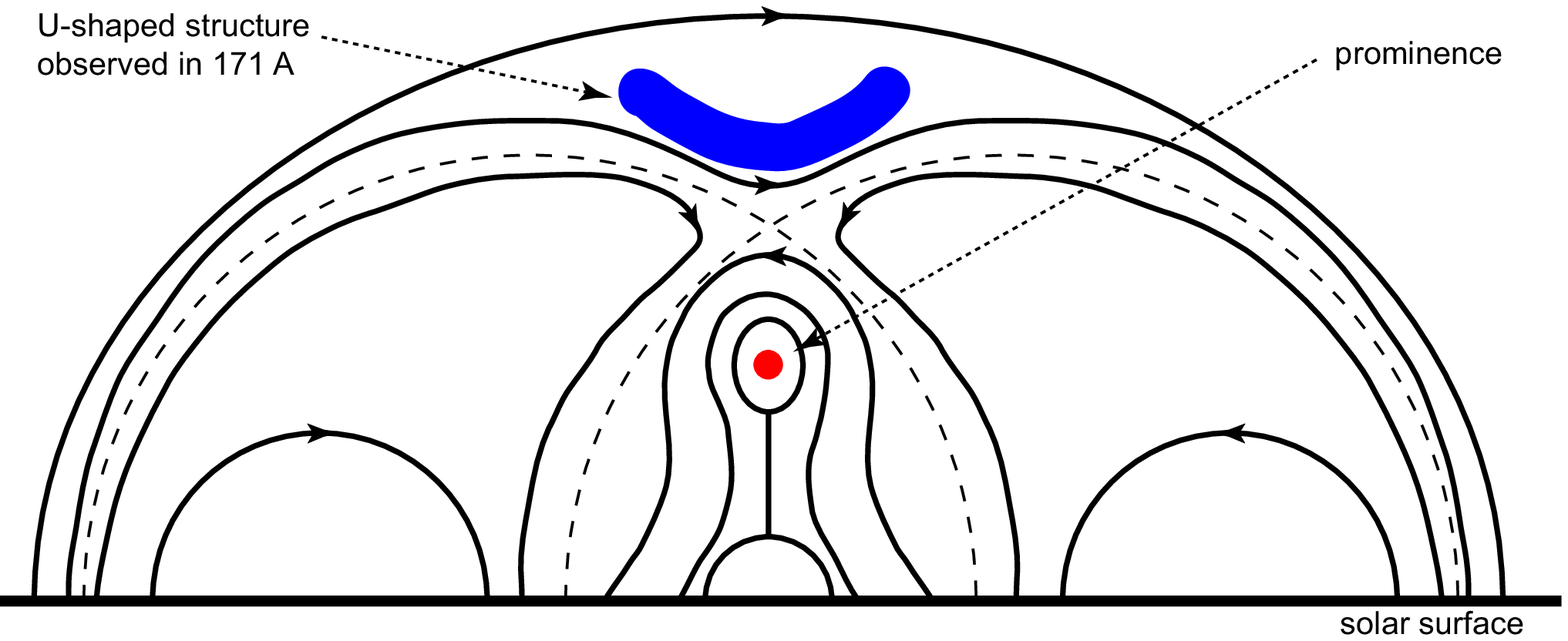}
\caption{CME breakout model and the observations.}
\label{F:Breakout}
\end{figure*}

Although the observations do not fit into the standard model, they fit into one of its modification---breakout model \citep{Antiochos1999}. In the breakout model the standard CME magnetic configuration is confined by the quadrupolar magnetic structure (see Figure~\ref{F:Breakout}). As the CME moves up, the overlying closed field lines reconnect and free the way for the CME. This model fits the observations, if we interpret the 171~\AA\ U-shaped structure as the field lines, which lie above the standard CME magnetic structure (see Figure~\ref{F:Breakout}). In this case the prominence will be below the U-shaped structure.

\subsection{Scattered Emission}

For the estimation of the CME mass, we assumed that the CME emission in the Fe~171~\AA\ is thermal (emitting ions are excited by collisions with thermal electrons). We neglected the contribution of the scattered emission because it is small in the low corona. But, as the CME goes away from the Sun, its electron density decreases, and the contribution of the scattered emission increases. At some distance, the scattered and thermal emissions could equalize, and the conclusion about the CME mass dynamics could be incorrect. In this section, we will estimate the contribution of the scattered emission to the CME intensity in the Fe~171~\AA\ line and its impact on the estimate of the CME mass.

In the coronal approximation, the number of the emitted photons in Fe~171~\AA\ line equals the number of the electrons excited to the upper level of the corresponding transition. The number of excitations due to the collision with thermal electrons is

\begin{equation}
\frac{dN_{th}}{dt} = N_{Fe}^{IX} n_e C_{coll},
\end{equation}
where $N_{Fe}^{IX}$---the number of Fe~IX ions; $n_e$---electron density; $C_{coll}$---collision strength of the transition. According to CHIANTI database \citep{Dere1997} for the transition corresponding to Fe~171~\AA\ line ($3s^23p^6$~$^1S_0$---$3s^23p^53d$~$^1P_1$), $C_{coll} = 1.36 \cdot 10^{-8}$~cm$^3$~s$^{-1}$.

The number of excitations due to the resonance scattering is
\begin{equation}
\frac{dN_{scatt}}{dt} = N_{Fe}^{IX} \sigma F,
\end{equation}
where $\sigma$---resonance scattering cross section; $F$---photons flux in the Fe~171~\AA\ line. 

We have calculated $F$ with the CHIANTI package: $F$~=~$10^{13}$~photons~s$^{-1}$~cm$^{-2}$. The $F$ decreases by a factor of 4 if the distance from the Sun's center increases from 1 to 2~$R_\odot$. For an estimation, we used the constant value of $F$ since the change in the electron density is much higher.

We calculated $\sigma$ with the formula \citep{syl86},

\begin{equation}
\sigma = \frac{\sqrt{\pi}e^2}{m_e c} f \lambda \sqrt{\frac{M_{Fe}}{2k_B T}} = \ 5 \cdot 10^{-16} \text{cm}^{-2}
\label{E:cross_section}
\end{equation}
where $e$---electron charge; $m_e$---electron mass; $c$---speed of light; $f$~=~$2.94$---transition oscillator strength; $\lambda$~=~171~\AA---transition wavelength; $k_B$---Boltzmann constant; $T$~=~1~MK---plasma temperature; $M_{Fe}$---ion Fe IX mass. 

For low corona electron density $n_e = 10^8$~cm$^{-3}$, the ratio of the scattered emission ($I_{scatt}$) to the thermal one ($I_{th}$) equals
\begin{equation}
\frac{I_{scatt}}{I_{th}} = \frac{dN_{scatt}/dt}{dN_{th}/dt} = \frac{\sigma F}{C_{coll} n_e} \sim 3 \cdot 10^{-3}
\end{equation}
This means that in the low corona, the CME emission is indeed thermal.

Let us estimate at what electron density the thermal and scattered emissions will equalize:
\begin{equation}
n_e \sim \frac{\sigma F}{C_{coll}} \sim 3 \cdot 10^5 \text{cm}^{-3}
\end{equation}
According to \citet{Vourlidas2010}, CMEs have such electron densities at a distance of 2~$R_\odot$ from the Sun's center. This means that our assumption about the thermal nature of the CME emission in the Fe~171~\AA\ line is violated only at the boundaries of the TESIS field of view, which does not influence our conclusion about the CME mass dynamics.

\subsection{CME Mass}

We considered two cases of CME geometry behavior: two-dimensional and three-dimensional. As previously mentioned, in reality we deal with some intermediate case, and the real mass dynamics is somewhere between these two cases. Since in both cases, the CME mass increased with time, we conclude that the real CME mass also increased with time. This implies that the CME beside the low corona plasma took away the plasma from the far corona.

However, the effect could be explained not only by the mass increase by accumulation of the far corona's plasma. First, the observed in the Fe~171~\AA\ line mass increase could be caused by the prominence heating. In this case, total CME mass is constant, but the CME mass is redistributing among hot and cool temperatures. This explanation does not contradict the observations: the prominence erupted, the prominence eventually faded away, and the prominence disappearance could occur due to its heating.

Second, we assumed that the CME temperature was constant. If initially the CME temperature was 1~MK or higher, and the CME temperature decreased with time, then the temperature would have moved towards the temperature response maximum of the TESIS Fe~171~\AA\ telescope (0.7~MK). This would have lead to an increase in CME intensity and therefore to higher values of the CME mass estimation.

Third, we did not calculate the CME mass; we estimated it. The CME mass increased by 30~--~150~\%; and therefore, the effect could be caused by the roughness of the model. So, although the estimations tell us that the CME mass increased with time, the effect could be caused by prominence heating, CME cooling, or roughness of the model. 

\subsection{Kinematics}

The main CME acceleration occurred at distances 2--6~$R_\odot$ from the Sun's center. This is consistent with other results on the CME kinematics, according to which the main acceleration occurs at distances 1.4--4.5~$R_\odot$ \citep{Zhang2001, Zhang2004, Gallagher2003}. 

After the CME erupted, the flare below A-level occurred. The lag of the flare behind the CME is common \citep{Zhang2001}. During the CME main acceleration phase, the flux in GOES 0.5--4~\AA\ slightly increased. We cannot confirm whether  this flux is related to the CME or not. If it is related to the CME, then it could be a sign of some reconnection processes taking place during the CME acceleration.

In the solar minimum, LASCO C2 observes CMEs in the ecliptic plane \citep{Yashiro2004}. However, erupting prominences associated with the CMEs in the solar minimum have bi-modal latitude distribution with maximums at the latitudes of $\pm30^{\circ}$ \citep{Plunkett2002}. It is considered, that this difference in the latitude distributions is caused by the presence of complicated multi-polar structures, which make the CME move along a curved trajectory \citep{Webb2012, Antiochos1999}. The results presented in this work show that CMEs indeed could move along a curved trajectory and that the high-latitude erupting prominence and the low-latitude LASCO CMEs could be connected with curved trajectory.

\section{Conclusion}

In this work, we presented results of observations of the CME, which occurred on May 13, 2009. The observations were carried out with TESIS EUV telescopes and LASCO coronagraphs. We observed the CME from the solar surface up to 15~$R_\odot$. The main results of the work are the prominence location, the CME mass dynamics, and the detailly measured kinematics. 

The CME was associated with an erupting prominence. We proved that the prominence was located in the lowest part of the U-shaped structure (observed in the Fe~171~\AA\ line) close to the magnetic X-point. This fact does not fit into the standard CME model, which expects the prominence to be above the U-shaped structure. We explained the observations with the CME breakout model. We interpreted the U-shaped structure as closed field lines, which lie above the standard CME magnetic structure.

We have estimated the CME mass for two different behaviors  of CME geometry. The CME mass had increased during its motion. Although the effect could be explained by other reasons---the prominence heating, the CME cooling, or the roughness of the model---we think that question of the CME mass dynamics deserves more attention.

Thanks to the large field of view of TESIS EUV telescopes, we observed the CME from the very beginning up to 15~$R_\odot$ without losing it from sight. Due to the relatively low CME velocity, we have a lot of the CME images. These two factors allowed us to measure the CME kinematics in detail. The CME kinematics is consistent with literature: three acceleration phases; distance range of the acceleration; the lag of the flare behind the CME. Moreover, we found out that the CME moved along a curved trajectory. Observation of the curved trajectory confirms the hypothesis of connection between high-latitude erupting prominences, observed with EIT~304~\AA, and low-latitudes CMEs, observed with LASCO C2.

The observations presented in this work show that EUV telescopes with wide fields of view are very effective in CME investigations. Thanks to the wide field of view, we can observe CMEs in the far corona, and thanks to the ability to observe in different wavelengths (for example in Fe~171~\AA\ and He~304~\AA\ lines), we can study how CME plasma with different temperatures interact with each other. We hope that in the future more ``EUV coronagraphs'' will be created and that they will shed some light on the nature of CMEs.

\acknowledgments
We are grateful to Ivan Loboda for his invaluable help. 

This work was  supported by a grant from the Russian Foundation of
Basic Research (grant 14-02-00945), and by the Program No. 22 for fundamental
research of the Presidium of the Russian Academy of Sciences.

\bibliographystyle{apj}
\bibliography{CME}
\end{document}